\pdfoutput=1
\RequirePackage{fix-cm}
\documentclass{article}
\usepackage{arxiv}

\usepackage{calc}
\usepackage{graphicx}
\usepackage{booktabs}
\usepackage{subcaption}
\usepackage{pdfpages}
\usepackage{amsmath}
\usepackage[inline]{asymptote}
\usepackage{bm}
\usepackage{algorithm}
\usepackage{algpseudocode}
\usepackage{hyperref}
\usepackage{siunitx}
\usepackage{tabularx}
\usepackage{xcolor}

\usepackage[T1]{fontenc}
\title{Application and Benchmark of SPH for Modeling the Impact in Thermal Spraying}

\author{Stefan Rhys Jeske\thanks{These authors contributed equally to this work.}\\
RWTH Aachen University\\
\texttt{jeske@cs.rwth-aachen.de}\\
         \And
        Jan Bender \\
RWTH Aachen University\\
\texttt{bender@cs.rwth-aachen.de}\\
        \And
        Kirsten Bobzin \\
RWTH Aachen University\\
\texttt{info@iot.rwth-aachen.de}\\
        \And
        Hendrik Heinemann \\
RWTH Aachen University\\
\texttt{heinemann@iot.rwth-aachen.de}\\
        \And
        Kevin Jasutyn$^*$ \\
RWTH Aachen University\\
\texttt{jasutyn@iot.rwth-aachen.de}\\
        \And
        Marek Simon \\
RWTH Aachen University\\
\texttt{simon@isf.rwth-aachen.de}\\
        \And
        Oleg Mokrov \\
RWTH Aachen University\\
\texttt{oleg.mokrov@isf.rwth-aachen.de}\\
        \And
        Rahul Sharma \\
RWTH Aachen University\\
\texttt{sharma@isf.rwth-aachen.de}\\
        \And
        Uwe Reisgen\\
RWTH Aachen University\\
\texttt{reisgen@isf.rwth-aachen.de}\\
}

\begin{document}

\maketitle

\begin{abstract}
The properties of a thermally sprayed coating, such as its durability or thermal conductivity depend on its microstructure, which is in turn directly related to the particle impact process.
To simulate this process we present a 3D Smoothed Particle Hydrodynamics (SPH) model, which represents the molten drop\-let as an incompressible fluid, while a semi-implicit En\-thalpy-Porosity method is applied for the mushy zone during solidification. 
In addition, we present an implicit correction for SPH simulations, based on well known approaches, from which we can observe improved performance and simulation stability. 
We apply our SPH method to the impact and solidification of Al$_2$O$_3$ drop\-lets onto a free slip substrate and perform a rigorous quantitative comparison of our method with the commercial software Ansys Fluent using the Volume of Fluid (VOF) approach, while taking identical physical effects into consideration. 
The results are evaluated in depth and we discuss the applicability of either method for the simulation of thermal spray deposition.
We show that SPH is an excellent method for solving this free surface problem accurately and efficiently.




\end{abstract}

\section{Introduction}
\label{sec:intro}
Thermal spraying is a coating technology where particles of a feedstock material are heated, fully or partially melted and accelerated to high speeds onto a substrate.
Through the impact of many particles a coating is built up. 
As a coating technology, thermal spraying is divided into three major process variants: flame spraying, electric arc spraying and plasma spraying \cite{Dav04}. 
Plasma spraying is characterized by particle velocities of up to v = \SI{800}{\metre\per \second} \cite{Paw08} and high plasma temperatures in the range of T = \SI{6000}{\degreeCelsius} to T = \SI{15000}{\degreeCelsius}, which are significantly above the melting temperature of any known material \cite{Dav04}. 
The injected particles are mostly in the size range of d = \SIrange{20}{90}{\micro\metre} \cite{Paw08}.
As the particles spread upon impact, rapid solidification occurs with cooling rates in the range of $\dot{q}$ = \SIrange{e7}{e8}{\kelvin\per\second} as a result of heat transfer from the liquid material to the underlying substrate and to the ambient atmosphere \cite{VMTC14}. 
Thus, the particle deformation on the substrate, cooling, and solidification occur simultaneously.
The properties of the coating, such as its durability or thermal conductivity are directly related to its microstructure, which is in turn directly related to the particle impact process. 
Therefore, a detailed understanding of the dynamics of particle impact on the substrate is essential for better control of the coating build-up. 
The deposition of particles during plasma spraying can be only poorly observed experimentally, due to the fact that splat formation and solidification occur in a few microseconds \cite{GF11}. 
Consequently, many studies have been devoted to numerical and analytical investigation of particle impact, splat formation, and solidification \cite{VVL+95,GMCC03,CF09,PPCM02,YLZD12}.

Besides the simulation of the plasma jet and the heat transfer from the plasma to the particles, the impact of the melted particle onto the substrate and its subsequent deformation are of great interest. 
A special challenge in modeling this process numerically is presented by the extreme dimensions, as the particles have a diameter of d = 20-90 \SI{}{\micro\metre} while traveling at speeds of v = \SI{200}{\metre\per\second}, as well as the rapid solidification process, where the thermal conditions decelerate the fluid motion, from melted to solid, on a very short time scale.

Given this context, we present our contribution as the construction of a novel SPH model which heavily utilizes implicit solvers to great effect, which improve both the simulation performance and stability. 
In this scope we also propose an implicit formulation for established SPH correction methods which aim to improve the physical accuracy.
In addition, we compare our SPH model against the commercial software Ansys Fluent by running simulations for a proposed benchmark model of droplet impact in thermal spray deposition.
From this we are able to show great agreement of the overall shape of the impacted droplet as well.
Using our SPH method, we were also able to simulate the droplet in significantly higher resolution while at the same time requiring a fraction of the computational cost.

\subsection{VOF-based Eulerian simulation approaches}
In the past, these impact simulations have often been performed with Eulerian methods using the volume of fluid (VOF) method. 
The VOF method can be used for the simulation of the free surface interface between two or more immiscible fluids by tracking the volume fraction of each of the fluids in two- or three-dimensional meshes. 
However, in the process of particle impact a large deformation of the molten particle occurs, from spherical to a thin layer.
Therefore, it requires a very fine, or spatially adaptive, mesh discretization over a large area of the simulation domain.
While fine meshes require exponentially increasing computational resour\-ces, mesh adaptivity is very difficult to implement correctly and still incurs a noticeable performance penalty.
Most of the computational cost of VOF algorithms is incurred by the cells that form the interfaces between different fluids \cite{BLC13}. 

Other than tracking the position of the free surface, the heat transfer and especially the effect of solidification strongly influences the dynamics of the particle impact process. 
A popular method to model the effect of solidification is the enthalpy-porosity method \cite{HA96}. 
Here, the heat transfer is solved in the enthalpy formulation, with addition of a source term that is dependent on the solid fraction of the semi-solidified fluid in the so called ``mushy zone'', to account for the latent heat of melting and solidification. 
Furthermore, another related source term is added to the momentum equation to account for increased flow resistance of the fluid in the semi-solidified region, due to growth of dendritic structures. 
This momentum sink, also called Darcy-term, is dependent on the permeability, which in turn is also dependent on the solid fraction in the ``mushy zone''.

Several works have applied the VOF method for modeling particle impact in the thermal spray process. 
Pasandideh-Fard et al.~\cite{PCM02} developed a 3D model to simulate the impact and solidification of a molten drop\-let on a flat substrate by applying the fixed velocity approach for solidification, where the solid is defined as liquid with infinite density and zero velocity. 
Another example is Zheng et al.~\cite{ZLZ+14}, who developed a 3D particle impact model during plasma spraying utilizing the momentum source method of Ansys Fluent for modeling the solidification. 

While these studies helped to increase the understanding of the particle impact and solidification in thermal spraying, a lot of computational time was generally required. 
In previous works of the authors the particle impact and solidification was modeled using a modified momentum source approach \cite{BOK+19} which was then applied to simulate multiple particle solidification \cite{BWH+21}. The most recent work presents a calculation of the effective thermal conductivity using information of inter-splat gaps derived from a simulation of multiple splats of Al$_{2}$O$_{3}$ droplets \cite{BWH21}. However, while the experimental results showed the characteristic length of the gaps to be <1 µm, the simulation could only resolve gaps with a width of its cell size of 2.25 µm.
This was resolved by numerically smearing the particle boundaries in this model and then enabling an approximately correct calculation of the thermal conductivity. However, this was only an interim solution, not a physically correct representation of the problem.

\subsection{Particle-based simulation approaches}

Due to the high computational cost of VOF-based Eulerian approaches, there has been a range of works applying the Smoothed Particle Hydrodynamics (SPH) method to the simulation of thermal spray deposition. 
Although the SPH method was originally introduced by Gingold and Monaghan \cite{GM77} and Lucy \cite{Luc77} in the field of Astrophysics, the method has already been used frequently for the simulation of the particle impact process.
This is often due to the versatility regarding the simulation of changing domain topology, free surfaces as well as multiple phases. 

Since the feedstock material in the thermal spray process consists of particles and at the same time the SPH method is based on discretization particles, first and foremost the terms should be properly distinguished. 
For this reason, the term \textit{particle} is from now on used for the SPH discretization particle of the numerical method, while the feedstock particle that is molten in the thermal spray process and projected towards the surface will be termed \textit{droplet}.

Fang et al.~\cite{FBW+09} introduce many ideas for the SPH simulation of droplet spreading and solidification.
They propose an improved pressure correction scheme and show simulations of droplets impacting on a solid substrate as well as similarities to some images obtained from experiments. 
In addition, for heat conduction an artificial heat model based on internal energy is used. 
Particles are classified into liquid, melting and solid, and a source term in the momentum equation accounts for phase change. 
Results are then compared against an experiment.

A similar approach is pursued by Zhang et al.~\cite{ZZZ08} without the pressure correction, but also considering melting of the substrate for high thermal conductivities, low thermal capacities and high droplet temperatures. However, a validation of the method was not performed.

Farrokhpanah et al.~\cite{FBM17,FMB21} present a novel method for the simulation of latent heat in SPH with specific application to Suspension Plasma Spraying, which is a variant of the thermal spray process.

Abubakar and Arif \cite{AA19} introduce a hybrid approach for the simulation of spray deposition. 
The SPH method is used to model the dynamics of splat formation during the spray process while the Finite Element Method (FEM) is used in order to model the solidification and compute residual stresses. 
The results are validated qualitatively by comp\-arison with experimental results in literature. 
While this approach is quite novel it relies on a very complex system in which numerical errors can occur at many different places (especially during transfers between discretizations). 
A coupled Eulerian and Lagrangian approach is also persued by Zhu et al.~\cite{ZKG15} where it is used in order to simulate spray deposition of semi-molten ceramic droplets.

While previous work has shown great progress in the simulation of thermal spray problems using SPH, most works are only partially validated against analytical solutions, e.g. Farrokhpanah et al.~\cite{FBM17,FMB21} validate heat conduction and solidification using analytical solutions, or against experimental data \cite{AA19,FBW+09,ZZZ08}.

In order to better understand the capabilities of the SPH method for modeling these kinds of problems we quantitatively compare the performance of our own implementation of the SPH method to the VOF method on Eulerian grids using Ansys Fluent for the specific conditions of the thermal spray process.

\section{Computational Method}
\label{sec:method}
\subsection{SPH Discretization}
\label{sec:method_sph}

In the following, we briefly outline our SPH discretization method of the Navier-Stokes equations for incompressible fluid flow.
In general, SPH is a Total-Lagrange\-ian discretization method which implies that fluid quantities are observed at positions which move along with the fluid.
These discrete positions, or particles, are advected and tracked through time and carry associated field quantities with them.
The SPH method uses a weighted interpolation, derived from the convolutional identity with the $\delta$-distribution, in order to compute unknown quantities and derivatives needed to solve Partial Differential Equations (PDEs).
An arbitrary scalar quantity $A_i = A(\bm{x}_i)$ at particle position $\bm{x}_i$ can be computed by weighted summation using
\begin{equation}
\label{eq:sph_sum}
A_i = \sum_{j\in \mathcal{N}_i} V_j A_j W(\bm{x}_i - \bm{x}_j; h),
\end{equation}
where $W(\bm{x}_i - \bm{x}_j; h)$ is a compactly supported weighting function, the commonly used cubic-spline kernel function in our case, around particle $i$ with smoothing length $h$ and $\sum_{j\in \mathcal{N}_i}$ denotes a summation over the neighboring particles $j$ of particle $i$, which lie within the compact support of $W$ centered on particle $i$.
Derivatives can easily be computed by differentiating Eq. \eqref{eq:sph_sum} which shifts the derivative operator to the weighting function. 
When doing this, however, care has to be taken since the commonly used cubic-spline kernel does not have a smooth second derivative.
For more information on derivatives, momentum conserving SPH sums, and SPH in general, the reader is referred to the works of Price \cite{Pri10} and Koschier et al.~\cite{KBST19}.

\subsubsection{Incompressible Fluid Model}
\label{sec:fluids}

The equations typically used for the simulation of incompressible fluids are the continuity equation, Eq. \eqref{eq:continuity}, and the Navier-Stokes equation, Eq. \eqref{eq:navier_stokes}:
\begin{align}
    \label{eq:continuity}
    \frac{D\rho}{Dt} &= 0 \quad \leftrightarrow \quad \frac{\partial \rho}{\partial t} = -\rho \nabla\cdot\bm{v}\\
    \label{eq:navier_stokes}
    \rho \frac{D\bm{v}}{D t} &= -\nabla p + \mu \nabla^2 \bm{v} + \bm{f}_{ext} + \bm{f}_{st}.
\end{align}
Here $\rho$ denotes the fluid density [kg m$^{-3}$], $\bm{v}$ the velocity [m s$^{-1}$], $p$ the pressure [N m$^{-2}$], $\mu$ the dynamic viscosity [Pa s], $\bm{f}_{ext}$ the external volumetric forces [N m$^{-3}$], e.g. gravity, and $\bm{f}_{st}$ the force due to surface tension [N m$^{-3}$]. 

\subsubsection{Pressure}

The pressure force is computed using the Divergence-Free SPH (DFSPH) method as presented by Bender and Koschier \cite{BK15}, which is an implicit solver ensuring both constant density and a divergence free velocity field, see Eq. \eqref{eq:continuity}.
We have found that this method allows us to use larger time steps during simulation, in contrast to explicit pressure solvers which compute the pressure force using an Equation of State (EOS), e.g. used by Farrokhpanah et al.~\cite{FMB21} and described by Monaghan \cite{Mon12}.
Also, recomputing the density (as is performed in DFSPH) in each time step avoids the possible loss of volume when advecting the local density using the continuity equation, Eq. \eqref{eq:continuity}.
While implicit pressure solvers have been explored in related works, e.g. see Fang et al.~\cite{FBW+09}, DFSPH also enforces a divergence free velocity field which has been shown by Bender and Koschier \cite{BK15} to improve the stability of the simulation.

\subsubsection{Viscosity}

The viscosity force is also computed implicitly using the model by Weiler et al.~\cite{WKBB18}.
It is obtained by solving for accelerations $\bm{a}_{visc}$ such that
\begin{equation}
    \label{eq:weiler_visko}
    \bm{a}_{visc} = \frac{\bm{v}_{visc}^{t+1} - \bm{v}^t}{\Delta t} = \nu\nabla^2 \bm{v}_{visc}^{t+1}.
\end{equation}
Discretizing this equation yields a system of linear equations for the velocity $\bm{v}_{visc}^{t+1}$ which is then used to compute the resulting acceleration due to viscous forces $\bm{a}_{visc}$ using the finite difference formula in Eq. \eqref{eq:weiler_visko}.
In general, the extensive use of implicit solvers enables the usage of larger simulation time steps without causing instabilities.

\subsubsection{Surface Tension}
\label{sec:surfacetension}

Surface tension computation in SPH is known to be a challenging problem, since it is very difficult to obtain a clear definition of the fluid surface.
There exist formulations based on fluid surface curvature, often derived from the Continuum Surface Force (CSF) model by Brackbill et al.~\cite{BKZ92}, as well as formulations based on inter-molecular forces.
For our purposes we have implemented the CSF model of Müller et al.~\cite{MCG03} based on the model of Morris \cite{Mor00}, where force $\bm{f}_{i, \text{st}}$, the curvature $\nabla^2 c_i$ and surface normal $\bm{n}_i$ are computed from a smoothed color field $c_i$:
\begin{align}
    c_i &= \sum_j \frac{m_j}{\rho_j} W_{ij}, \\ 
    \bm{n}_i &= \sum_j \frac{m_j}{\rho_j} (c_j - c_i) \nabla W_{ij}, \\  
    \nabla^2 c_i &= -\sum_j \frac{m_j}{\rho_j} (c_i - c_j) \frac{2 ||\nabla W_{ij}||}{||\bm{x}_i - \bm{x}_j|| + \varepsilon }, \\
    \bm{f}_{i, \text{st}} &= -\sigma \nabla^2 c_i \frac{\bm{n}_i}{||\bm{n}_i||}.
    \label{eq:surface_tension}
\end{align}
Here $\sigma$ denotes the surface tension coefficient [\si{\N\per\m}].
The computation of the normals and especially the curvature is documented to be prone to errors due to particle disorder, however we have not observed any significant instabilities in our simulations.
This could be due to the extremely small time scale of our simulations as well as due to other forces being more dominant.

\subsubsection{Solidification}
\label{sec:solidification}
Solidification is one of the main determining factors of the dynamics of the thermal spray process. 
It depends on the splat thickness, the thermal diffusivities of both the sprayed feedstock material as well as the underlying solid material, and the thermal contact resistance between the flattening droplet and the substrate. 
It directly affects the deformation behavior, the splat shape and the coating microstructure \cite{Paw08}. 
However, as the SPH particle radius had to be chosen very small in order to accurately capture the dynamics of the droplet impact, the substrate was not included in the calculation domain but instead modeled as a solid wall (boundary condition) with constant temperature $T=$ \SI{300}{\kelvin} and the thermal contact resistance is therefore only determined by the thermal conductivity. 
Hence, the fluid of the droplet is cooled upon contact with the wall and the subsequent solidification process is modeled by taking into account a Darcy-term (momentum sink), after the well known Enthalpy-Porosity method, for modeling the solidification of pure metals \cite{BVR88} and of binary alloys \cite{VBP90}. 
While the latent heat of melting and solidification was not yet considered in the heat transfer equations of our model, Eq. \eqref{eq:energy_transport}, the formulation in terms of the enthalpy Eq. \eqref{eq:enthalpy_temperature_heat_transfer} allows for the consideration of latent heat with little modification in the future. 
This simplification was chosen to allow for a more simple and consistent first comparison of the numerical methods, with the option of including the non-linearities of the material parameters in subsequent works. 

The considered Darcy-term adds an acceleration to the Navier-Stokes equation which has a strong movement inhibiting effect on the fluid, once the temperature of the fluid becomes low enough. 
This so called momentum sink accounts for the semi-liquid state in the so called mushy-zone, where already some nucleation and dendrite growth has occurred, thereby affecting the properties of the fluid. 
The effect is controlled by the liquid fraction $f_l(T)$, here modeled as a simple Heaviside function, Eq. \eqref{eq:heaviside_liquidfraction}, and by a morphological constant $C$:
\begin{align}
\bm{a}_{porosity} &= -\bm{v} C f_l(T) \\
f_l(T) &= \begin{cases}
0 & T > T_l\\
1 & T_l-\Delta T_l \leq T \leq T_l\\
- &   T_l \leq T-\Delta T_l. 
\end{cases}
\label{eq:heaviside_liquidfraction}
\end{align}
In this equation $C$ has the unit [s$^{-1}$] which, intuitively, is related to the time span required during which the fluid will solidify completely, given no other influences.
In order to be able to capture the solidification process, regardless of simulation method, the maximum time step of the simulation should be selected to be smaller than $C^{-1}$.
The values for $C$ are often very large, resulting in very large deceleration as soon as $T\leq T_l$, so large in fact that simulations using explicit time stepping can become unstable.
These instabilities are a result of the material solidifying in less than a single simulation step.
This is remedied by constructing an algebraic equation which computes the acceleration using the projected velocity of the next time step as in the following
\begin{align}
    \bm{a}_{porosity} = \frac{\bm{v}^{t+1} - \bm{v}^{t}}{\Delta t} &= -\bm{v}^{t+1} C f_l(T)\\
    \bm{v}^{t+1} &= \bm{v}^t \frac{1}{1 + \Delta t C f_l(T)}.
\end{align}
The acceleration is then simply computed by inserting the expression for $\bm{v}^{t+1}$
\begin{equation}
    \label{eq:momentum_sink}
    \bm{a}_{porosity} = \frac{\bm{v}^t}{\Delta t}\left( \frac{1}{1 + \Delta t C f_l(T)} - 1\right).
\end{equation}
This semi-implicit formulation allows the usage of larger time steps in the simulation without causing instabilities, but comes at the cost of slightly dampening the effect.
Additionally, if the time step is very large, the solidification may occur very quickly.
Nevertheless, since the observed time intervals are often too large to observe the solidification process of single particles anyway, this is deemed to be an acceptable trade-off.
The parameters used for the momentum-sink are shown in Table \ref{tab:momentumsink}.

\subsubsection{Correction Terms}
\label{sec:correction}

In our implementation we have found that it is also necessary to add correction terms, which improve the quality of SPH simulations.
The first such term was documented by Monaghan \cite{Mon89}, and reduces the interpenetration of particles by smoothing the velocity field while conserving linear and angular momentum, without adding dissipation.
The acceleration of this correction, also sometimes called XSPH viscosity, is given by
\begin{equation}
\label{eq:xsph}
    \bm{a}_{i, xsph} = - \frac{\alpha}{\Delta t} \sum_{j\in\mathcal{N}_i} \frac{\overline{m}_{ij}}{\overline{\rho}_{ij}} (\bm{v}_i - \bm{v}_j) W_{ij},
\end{equation}
where $\alpha$ denotes the (dimensionless) strength of this smoothing, $\overline{m_{ij}}$ the average mass between particle $i$ and $j$ and $\overline{\rho_{ij}}$ the averaged density.
In the original work $\alpha = 1$ was proposed, yet we have found smaller values to also work very well.

The second correction addresses the issue of tensile instability at the surface of SPH fluids.
When using a method which recomputes the density instead of advecting it, it occurs that the density estimate at free surfaces is erroneous due to missing particles and causes an uncontrollable artificial surface tension effect.
This is solved in the DFSPH method by only considering ``over''-pressures due to larger density values and clamping smaller densities to the rest density.
This entirely removes the instability at the surface, but comes at the cost of reduced, non-surface tension, fluid cohesion.
In order to restore fluid cohesion a corrective force of the form
\begin{equation}
\label{eq:cohesion}
    \bm{a}_{i,cohesion} = -\gamma \sum_{j\in\mathcal{N}_i} \frac{\overline{m}_{ij}}{\overline{\rho}_{ij}} (x_i - x_j) W_{ij},
\end{equation}
is employed, where $\gamma$ is a parameter controlling the strength of cohesion [\si{\N\per\m}].
By design this correction term only adds forces in regions with particle deficiency and is inspired by the work of Monaghan \cite{Mon00}, yet instead of adding an additional repulsion term the attraction is first clamped and then reintroduced.
A very similar cohesive force was also used by Becker and Teschner \cite{BT07}.

We note that due to the anti-symmetric nature of these formulations, both conserve angular and linear momentum.
In order to further improve the stability and in order to be able to use larger time steps, we formulate these corrections implicitly in terms of velocity and incorporate them together with the viscous force into a single linear system
\begin{equation}
\label{eq:implicit_system}
\begin{split}
    \frac{\bm{v}_i^{t+1} - \bm{v}_i^t}{\Delta t} =& \frac{\mu}{\rho_i} 2(d+2) \sum_{j\in\mathcal{N}_i}  \frac{\overline{m}_{ij}}{\rho_j} \frac{\bm{v}_{ij}^{t+1}\cdot\bm{x}_{ij}}{||\bm{x}_{ij}||^2 + 0.01 h^2} \nabla W_{ij} \\
    &- \frac{\alpha}{\Delta t} \sum_{j\in\mathcal{N}_i} \frac{\overline{m}_{ij}}{\overline{\rho}_{ij}} \bm{v}_{ij}^{t+1} W_{ij} \\
    &- \gamma \sum_{j\in\mathcal{N}_i} \frac{\overline{m}_{ij}}{\overline{\rho}_{ij}} (\bm{x}_{ij}^{t} + \Delta t\bm{v}_{ij}^{t+1}) W_{ij},
\end{split}
\end{equation}
where $\bm{v}_{ij} = \bm{v}_i - \bm{v}_j$, $\bm{x}_{ij} = \bm{x}_i - \bm{x}_j$ and d is the number of spatial dimensions, i.e. $d=3$.
This linear system is solved using the matrix-free conjugate gradient method.

We are not aware of any other works utilizing implicit solvers in SPH simulations to this degree.
Doing this we are able to observe a very significant performance improvement due to the ability of simulating with large time steps without causing instabilities.

Pseudocode of our full fluid solver is shown in Algorithm \ref{lst:fluid_solver}.
It can be seen that the only explicitly computed component is the surface tension, while the pressure and remaining forces are independently implicitly integrated.

\begin{algorithm}[t]
  \caption{Fluid Solver}\label{lst:fluid_solver}
  \begin{algorithmic}[1]
    \Procedure{FluidSolve}{}
    \State\Call{SolveDivergenceFreeVelocity}{$i$}\Comment{\cite{BK15}}
    \State\Call{SolveViscosityAndCorrections}{$i$}\Comment{Eq. \eqref{eq:implicit_system}}
    \For{all particles $i$}
      \State\Call{ComputeSurfaceTension}{$i$}\Comment{Eq. \eqref{eq:surface_tension}}
      \State\Call{SolveMomentumSink}{$i$}\Comment{Eq. \eqref{eq:momentum_sink}}
    \EndFor
    \State\Call{SolveConstantDensityPressure}{$i$}\Comment{\cite{BK15}}
    \State $\Delta t \gets$ \Call{ComputeCFLTimeStep}{\,}
    \For{all particles $i$}
      \State $\bm{v}_i \gets \bm{v}_i + \Delta t \bm{a}_i$
      \State $\bm{x}_i \gets \bm{x}_i + \Delta t \bm{v}_i$
    \EndFor
    \EndProcedure
  \end{algorithmic}
\end{algorithm}

\subsection{Heat Transfer}
\label{sec:method_heat_transfer}

Heat conduction is governed by the Fourier equation 
\begin{equation}
    \frac{D(\rho c_p T)}{D t} =  \nabla\cdot\left(\lambda \nabla T \right) + \dot{q}''',
    \label{eq:energy_transport}
\end{equation}
where $\rho$ denotes the material density {[}kg m$^{-3}${]}, $c_p$ the specific heat capacity {[}J kg$^{-1}$ K$^{-1}${]}, $T$ the temperature {[}K{]}, $\lambda$ the thermal conductivity {[}W K$^{-1}$ m$^{-1}${]} and $\dot{q}'''$ the contribution from volumetric heat sources {[}W m$^{-3}${]}. 
In all of our simulations we set $\dot{q}''' = 0$, since we do not need any external heat sources.
Instead of using the temperature as the main variable for heat transfer, we transform Eq. \eqref{eq:energy_transport} using the relationship between specific enthalpy $h$ {[}J kg$^{-1}${]} and the temperature $T$
\begin{equation}
    h(T) = \int_0^T c_p(T) dT,
\label{eq:enthalpy_temperature}
\end{equation}
where $c_p$ may also be a function of temperature, taking into account e.g. the latent heat of melting. 
Assuming that $c_p(T)$ is continuous results in a bijective function, such that $h(T)$ as well as $T(h)$ are well defined.
This results in the following equation which uses both the specific enthalpy $h$ as well as the temperature:
\begin{equation}
    \rho\frac{D(h)}{D t} =  \nabla\cdot\left(\lambda \nabla T \right).
    \label{eq:enthalpy_temperature_heat_transfer}
\end{equation}
The change in density $\frac{D\rho}{Dt} = 0$ is already enforced by the constant density component of the implicit pressure solver. 
The equation above is discretized using SPH and explicit Euler time integration, resulting in the following discrete equation for the fluid particle with index $i$
\begin{equation}
    \rho_i \frac{h_i^{t+1} - h_i^t}{\Delta t} = \nabla \cdot (\lambda \nabla T)_i^t.
    \label{eq:heat_conduction_integration}
\end{equation}
The discretization of the heat conduction term is given in the following equation
\begin{equation}
    \label{eq:heat_conduction}
    \nabla \cdot (\lambda \nabla T)_i = \sum_{j\in \mathcal{N}_i}\frac{m_j}{ \rho_j}\frac{4\lambda_i\lambda_j}{\lambda_i + \lambda_j}(T_i - T_j)\frac{ \nabla_i W_{ij}\cdot \bm{r}_{ij}}{||\bm{r}_{ij}||^2},
\end{equation}
as is also the case for other related work, e.g. Zhang et al.~\cite{ZZZ08}, and was initially proposed by Brookshaw \cite{Bro85}.
It should be noted that $\lambda_i = \lambda(T_i)$ is generally a function of temperature and that $\bm{r}_{ij} = \bm{x}_i - \bm{x}_j$ is the vector between the positions of particle $i$ and particle $j$.

Since heat can only be conducted within the material itself, the SPH formulation is adiabatic by construction.
Finally, in order to reduce the computational requirements during simulation, the enthalpy is precomputed in terms of the temperature by integration of Eq. \eqref{eq:enthalpy_temperature}. 
Our heat solver algorithm is outlined in Algorithm \ref{lst:compute_heat_transfer} while the overall simulation pipeline is summarized in Algorithm \ref{lst:sph_solver}. 
\begin{algorithm}[t]
  \caption{Heat Transfer Computation}\label{lst:compute_heat_transfer}
  \begin{algorithmic}[1]
    \Procedure{HeatSolve}{}
    \For{all particles $i$}
      \State\Call{ComputeTemperature}{$i$}\Comment{Eq. \eqref{eq:enthalpy_temperature}}
    \EndFor
    \For{all particles $i$}
      \State\Call{ComputeHeatConduction}{$i$}\Comment{Eq. \eqref{eq:heat_conduction}}
    \EndFor
    \For{all particles $i$}
      \State\Call{ExplicitEulerIntegration}{$i$}\Comment{Eq. \eqref{eq:heat_conduction_integration}}
    \EndFor
    \EndProcedure
  \end{algorithmic}
\end{algorithm}

\begin{algorithm}[t]
  \caption{Solver Overview}\label{lst:sph_solver}
  \begin{algorithmic}[1]
    \Procedure{FluidAndHeatSolver}{}
    \State $t \gets t_\text{start}$
    \While{$t < t_\text{end}$}
      \State\Call{HeatSolve}{\,}\Comment{Algorithm \ref{lst:compute_heat_transfer}}
      \State\Call{FluidSolve}{\,}\Comment{Algorithm \ref{lst:fluid_solver}}
      \State $t \gets t + \Delta t$
    \EndWhile
    \EndProcedure
  \end{algorithmic}
\end{algorithm}

\subsection{Boundary Conditions}

Boundary contributions for all SPH terms are computed using the approach of Akinci et al.~\cite{AIA+12}.
The only boundary present in our simulation is the substrate, which will be described further in Section \ref{sec:method_domain}.
Using the approach of Akinci et al. the boundary is sampled using a single layer of particles on the surface of the boundary.
The contribution of the boundary to the SPH summation of fluid particles can be generalized as
\begin{equation}
    A_i = A_i^f + A_i^b = \sum_{j \in\mathcal{N}_i^f} V_j A_j W_{ij} + \sum_{j \in\mathcal{N}_i^b} V_j^b  A_j^b W_{ij}.
\end{equation}
The superscript $f$ indicates contribution from fluid particles while $b$ indicates contribution from boundary particles within the compact support of particle $i$.
The volume of boundary particles is computed using
\begin{equation}
    V_i^b = \frac{1}{\sum_{j\in\mathcal{N}_i^b} W_{ij}},
\end{equation}
which is an SPH summation over the other boundary particles in the compact support of boundary particle $i$.
The boundary volumes are incorporated into the summations of fluid particles by extending fluid quantities into the boundary region.
This means e.g. using the rest density of the fluid to compute a mass from the boundary volume and utilizing this contribution for fluid density computations.
Similar considerations can be made for all other cases, such as for the Fourier equation.
The heat conduction term in Eq. \eqref{eq:heat_conduction} is extended by the following term for the boundary contribution
\begin{equation}
    \nabla \cdot (\lambda \nabla T)_i^b = \sum_{j\in \mathcal{N}_i^b} V_j^b 2\lambda_i (T_i - T_j^b)\frac{ \nabla_i W_{ij}\cdot \bm{r}_{ij}}{\bm{r}_{ij}^2}.
\end{equation}
This is equivalent to Eq. \eqref{eq:heat_conduction}, when also using $\lambda_j^b = \lambda_i$ and prescribing the wall temperature $T_j^b$.
This allows specifying a Dirichlet boundary condition on the substrate.

\subsection{Simulation Domain}
\label{sec:method_domain}

\begin{figure}
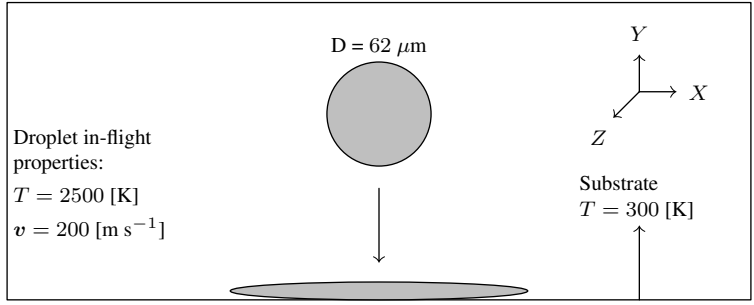

\centering
\begin{asy}[width=.6\linewidth]
usepackage("bm");
settings.outformat = "pdf";
defaultpen(fontsize(8pt));
// Box
draw(box((-0.5, 0), (0.5, 0.4)));
// Particle
filldraw(circle((0, 0.25), 0.07), mediumgray);
label("D = $62$ $\mu$m", (0, 0.32), N);
draw((0, 0.15) -- (0, 0.05), arrow=Arrow(TeXHead));
filldraw(ellipse((0, 0.012), 0.2, 0.012), mediumgray);
// Properties
label(minipage("Droplet in-flight\\
\smallskip
properties:\\
\smallskip
$T = 2500$ [K]\\
\smallskip
$\bm{v} = 200$ [m s$^{-1}$]", 60), (-0.5, 0.15), E);
// Substrate
draw((0.35, 0) -- (0.35, 0.1), arrow=Arrow(TeXHead));
label(minipage("\raggedright Substrate\\
$T = 300$ [K]", 45), (0.35, 0.1), N);
// Coordinate
draw((0.35, 0.28) -- (0.35, 0.33), arrow=Arrow(TeXHead));
draw((0.35, 0.28) -- (0.40, 0.28), arrow=Arrow(TeXHead));
draw((0.35, 0.28) -- (0.315, 0.245), arrow=Arrow(TeXHead));
label("$X$", (0.43, 0.26), N);
label("$Y$", (0.35, 0.34), N);
label("$Z$", (0.295, 0.200), N);
\end{asy}
\caption{Schematic diagram of the simulation domain for the droplet impact.}
\label{fig:schema_particle_impact}
\end{figure}

In the previous sections we introduced our simulation model for droplet impact onto a substrate with solidification.
Subsequently, we describe the simulation setup used to compare our SPH model with Ansys Fluent.
Figure \ref{fig:schema_particle_impact} shows the simulation domain for droplet impact. 
This includes the material properties, the initial droplet in-flight properties, the substrate wall properties and boundary conditions. 
The material properties of the Al$_{2}$O$_{3}$ droplet are listed in Table \ref{tab:material_properties_al2o3}. Further, the simulation parameters in Ansys Fluent and for SPH are listed in Table \ref{tab:simulationparameters}.

\begin{table}[b]
\begin{subtable}[h]{\linewidth}
    \centering
    \begin{tabular}{lll}
        \noalign{\smallskip}\hline\noalign{\smallskip}
        Property & Unit                      & Value        \\ \noalign{\smallskip}\hline\noalign{\smallskip}
        Droplet diameter & \SI{}{\micro\metre}     & 62  \\
        Density & kg m$^{-3}$               & 3,950  \\
        Specific heat  & J kg$^{-1}$ K$^{-1}$      & 900   \\
        Thermal conductivity    & W m$^{-1}$ K$^{-1}$       & 30    \\
        Viscosity & kg m$^{-1}$ s$^{-1}$      & 0.055 \\
        Surface tension & kg s$^{-2}$ & 0.8 \\
        Melting temperature     & K                         & 2,027  \\
        \noalign{\smallskip}\hline
    \end{tabular}
    \caption{Material properties Al$_2$O$_3$ \cite{BWH21}}
    \label{tab:mat_properties}
\end{subtable}
\begin{subtable}[h]{\linewidth}
    \centering
    \begin{tabular}{lll}
        \noalign{\smallskip}\hline\noalign{\smallskip}
        Property & Unit & Value        \\ \noalign{\smallskip}\hline\noalign{\smallskip}
        Wall temperature & K    & 300   \\
        Free-slip boundary condition\\
        \noalign{\smallskip}\hline
    \end{tabular}
    \caption{Wall boundary conditions}
    \label{tab:wall_boundary_conditions}
\end{subtable}
\caption{Material properties of Al$_2$O$_3$ droplet and boundary conditions.}
\label{tab:material_properties_al2o3}
\end{table}

\begin{table}[t]
    \begin{subtable}[h]{\linewidth}
        \centering
        \begin{tabular}{lll}
        \hline\noalign{\smallskip}
        Property                & Unit      & Value        \\ \noalign{\smallskip}\hline\noalign{\smallskip}
        Particle radius         & \SI{}{\micro\metre}        & 0.4  \\
        CFL min time step size  & s         & 1$\times 10^{-12}$    \\
        CFL max time step size  & s         & 1$\times 10^{-8}$  \\
        Simulation method       & -         & DFSPH \\
        XSPH correction $\alpha$ Eq.\eqref{eq:xsph} & - & 0.3  \\
        Cohesion correction $\gamma$ Eq.\eqref{eq:cohesion}& \SI{}{\newton\per\metre}  & 200  \\
        \noalign{\smallskip}\hline 
        \end{tabular}
       \caption{Simulation parameters in SPH}
       \vspace{0.5 cm}
\label{tab:simulationsparameter_SPH} 
    \end{subtable}
    
    \begin{subtable}[h]{\linewidth}
    \centering
    \begin{tabular}{lll}
    \hline\noalign{\smallskip}
    Property                & Unit          & Value        \\ \noalign{\smallskip}\hline\noalign{\smallskip}
    Domain size             & \SI{}{\micro\metre} & 225$\times$225  \\
    Mesh size               & \SI{}{\micro\metre} & 2.25  \\
    CFL min time step size  & s             & 1$\times$$10^{-12}$  \\
    CFL max time step size  & s             & 1$\times$$10^{-7}$  \\
    Simulation method       & -             & FVM  \\
    \noalign{\smallskip}\hline 
    \end{tabular}
       \caption{Simulation parameters in Ansys Fluent}
       \vspace{0.5 cm}
\label{tab:simulationsparameter_Fluent} 
    \end{subtable}
    
    \begin{subtable}[h]{\linewidth}
        \centering
        \begin{tabular}{lll}
        \hline\noalign{\smallskip}
        Property         & Unit      & Value        \\ \noalign{\smallskip}\hline\noalign{\smallskip}
        $T_l$            & K         & 2,027  \\
        $\Delta T_l$     & K         & 100   \\
        $C$              & s$^{-1}$  & 3$\times 10^{-7}$  \\
        \noalign{\smallskip}\hline 
    \end{tabular}
        \caption{Momentum sink parameters}
\label{tab:momentumsink}
     \end{subtable}
\caption{Simulation parameters}
\label{tab:simulationparameters}
\end{table}

\subsubsection{Ansys Fluent}

A 3D thermal spray coating build-up model based on a previous publication of the authors \cite{BWH21}, was created and implemented in Ansys Fluent. 
In this model the impact of thermally sprayed Al$_{2}$O$_{3}$ droplets onto a flat substrate was simulated. 
A momentum source function was used to simplify the calculation of the solidification process, with parameters as seen in Table \ref{tab:momentumsink}. 
A laminar viscous model was used and the energy solver was enabled. 
The dimensions of the spatial domain are 225 µm x 225 µm x 75 µm, incorporating the droplet as well as a surrounding gaseous atmosphere which was assumed to be air.
To shorten the computation time for the simulation of the impact of multiple droplets, a mesh edge length of 2.25 µm was chosen.
The calculation mesh consists of 330,000 cells.
The boundaries of the domain consists of an inlet for droplet on top, with velocity $\bm{v}_{p}=$ \SI{200}{\metre\per\second}, the substrate as a free-slip wall at the bottom and outlets with pressure $p_{amb}=101,325$ Pa and backflow total temperature $T_{out}=3,000$ K.
The interior domain is filled with air at $T_{gas}=2,400$ K at the start of the simulation. 
The temperature of the substrate is set to $T_{substrate} = 300$ K.
The free-slip boundary condition was applied for the contact between the droplet and the substrate and a VOF-approach was assumed for the calculation of the free surface of the Al$_{2}$O$_{3}$ droplet and the surrounding gas phase. 
The numerical parameters are summarized in Table \ref{tab:simulationsparameter_Fluent}.
The simulation was performed using Ansys Fluent 2020 R2 and computed on 32 cores of a compute cluster. The computation time was approximately 20 minutes.

\subsubsection{SPH}

\begin{figure}[t]
\centering
  \def\svgwidth{.5\linewidth}
\begingroup%
  \makeatletter%
  \providecommand\color[2][]{%
    \errmessage{(Inkscape) Color is used for the text in Inkscape, but the package 'color.sty' is not loaded}%
    \renewcommand\color[2][]{}%
  }%
  \providecommand\transparent[1]{%
    \errmessage{(Inkscape) Transparency is used (non-zero) for the text in Inkscape, but the package 'transparent.sty' is not loaded}%
    \renewcommand\transparent[1]{}%
  }%
  \providecommand\rotatebox[2]{#2}%
  \newcommand*\fsize{\dimexpr\f@size pt\relax}%
  \newcommand*\lineheight[1]{\fontsize{\fsize}{#1\fsize}\selectfont}%
  \ifx\svgwidth\undefined%
    \setlength{\unitlength}{173.38165267bp}%
    \ifx\svgscale\undefined%
      \relax%
    \else%
      \setlength{\unitlength}{\unitlength * \real{\svgscale}}%
    \fi%
  \else%
    \setlength{\unitlength}{\svgwidth}%
  \fi%
  \global\let\svgwidth\undefined%
  \global\let\svgscale\undefined%
  \makeatother%
\vspace{10 mm}
  \begin{picture}(1,0.56104421)%
    \lineheight{1}%
    \setlength\tabcolsep{0pt}%
    \put(0,0){\includegraphics[width=\unitlength]{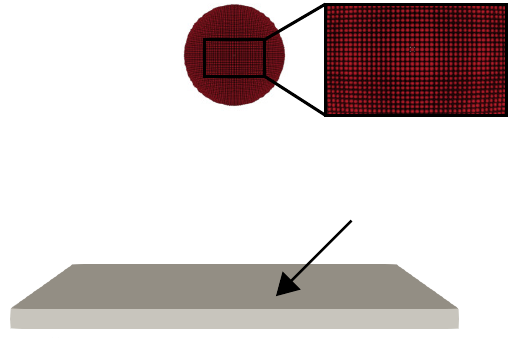}}%
    \put(0.02365547,0.32392956){\color[rgb]{0,0,0}\makebox(0,0)[lt]{\lineheight{1.25}\smash{\begin{tabular}[t]{l}Droplet diameter: 62 µm\\Particle radius: 0.4 µm\\A droplet consists of \\238,310 particles\end{tabular}}}}%
    \put(0.5162526,0.28693047){\color[rgb]{0,0,0}\makebox(0,0)[lt]{\lineheight{1.25}\smash{\begin{tabular}[t]{l}Wall $T = 300$ K\\Free-slip boundary condition\end{tabular}}}}%
  \end{picture}%
\endgroup%

\caption{Computational domain of droplet impact simulation in SPH.}
\label{fig:domain_SPH}
\end{figure}

The same 3D thermal spray coating build-up model was created using our SPH method. 
The domain of the simulation model is shown in Figure \ref{fig:domain_SPH}. 
The numerical parameters for the simulation and for the momentum sink are each listed in Table \ref{tab:simulationsparameter_SPH} and Table \ref{tab:momentumsink}, respectively. 
The droplet has a diameter of $d =$ \SI{62}{\micro\metre} and initial in-flight properties such as temperature of $T = 2,500$ K and velocity of v = \SI{200}{\metre\per\second}.
The droplet was discretized with particles with an individual radius of $r =$ \SI{0.4}{\micro\metre} and consists of 238,310 particles.
The substrate was modeled as a rigid body with a Dirichlet boundary condition for the temperature of $T_{wall} = 300$ K and a free-slip boundary condition for the momentum equation.
The model was implemented in a custom branch of SPlisHSPlasH \cite{splishsplash} and computed on 32 cores of the same compute cluster.
The computation time was approximately 5 minutes.

\section{Results}
\label{sec:results}
\begin{figure}[t]
\vspace{5 mm}
\centering
\def\svgwidth{.5\linewidth}
  \input{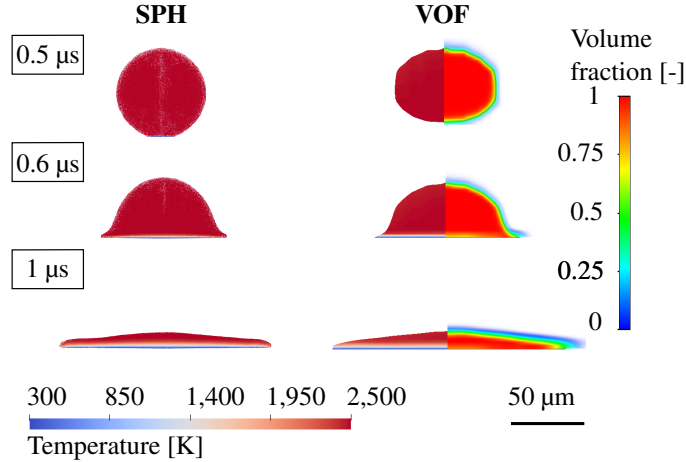}
\caption{Comparison of cross-sections of the droplet impact in SPH (left) and Ansys Fluent/VOF=0.5 (right), for several points in time.}
\label{fig:particle_impact_vergleich}
\end{figure}

\begin{figure}[b]
  \centering
\def\svgwidth{.5\linewidth}
  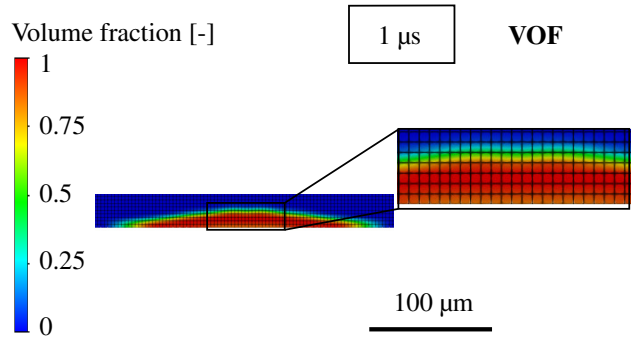
\caption{Cross section of volume fraction of the ceramic phase in Ansys Fluent at $t$ = \SI{1.0}{\micro\second}.}
\label{fig:vof_cross_section}
\end{figure}

\begin{figure}
  \centering
\def\svgwidth{.5\linewidth}
  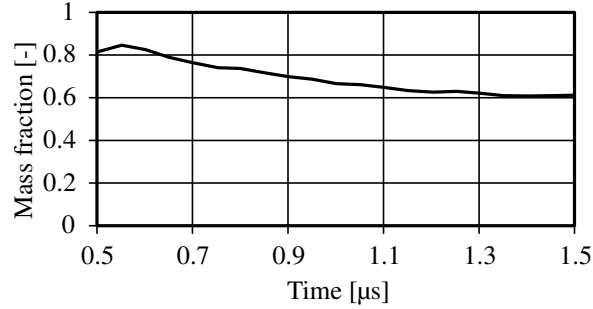
\caption{Mass share of the ceramic phase enclosed within the volume VOF$\geq 0.5$ of the total mass of the ceramic phase.}
\label{fig:volume_loss}
\end{figure}

Figure \ref{fig:particle_impact_vergleich} shows the droplet impact and the subsequent spreading of the droplet and solidification process modeled with SPH (left) and Ansys Fluent (right). 
The main dynamics of the process occurs at the shown three points in time.
It can be seen that there is a relatively high agreement between both methods. 
However, it should be noted, that the Ansys Fluent approach is performed with the VOF method for the modeling of the free surface of the liquid and therefore the presented shape represents the iso-surface of volume fraction 0.5 of the ceramic phase. 
While the overall resolution of the mesh is quite high, the mesh is relatively coarse in the region of interest, as can be seen in Figure \ref{fig:vof_cross_section}.
As such, the dispersion of the fluid boundary surface can be considerable in the VOF method and the apparent area of the cross section appears somewhat smaller than the area of the cross section in the SPH method, although in both cases the total mass is conserved.
Nevertheless, Figure \ref{fig:volume_loss} shows the decrease of the ratio of mass enclosed by the VOF-0.5 contour with respect to the total fluid mass, as the mass disperses across a larger region.
After impact at $t$ = \SI{0.5}{\micro\second}, the mass share of the VOF-0.5 contour of the droplet decreases steadily and then levels out at $t$ = \SI{1.4}{\micro\second}.

Another peculiarity is the shape of the droplet in free flight. 
While in the SPH method, the shape is highly spherical, the VOF droplet appears to be elliptically compressed in the direction of flight in the VOF method. 
While the surrounding region of the droplet is filled with stagnant air, the droplet itself is immersed in an airstream of the impact velocity in order to avoid compression due to drag.
As such, the slightly compressed shape can be explained by difficulties of achieving a perfect droplet shape using a transient inlet function.

\begin{figure}[b]
\centering
\def\svgwidth{.5\linewidth}
  \input{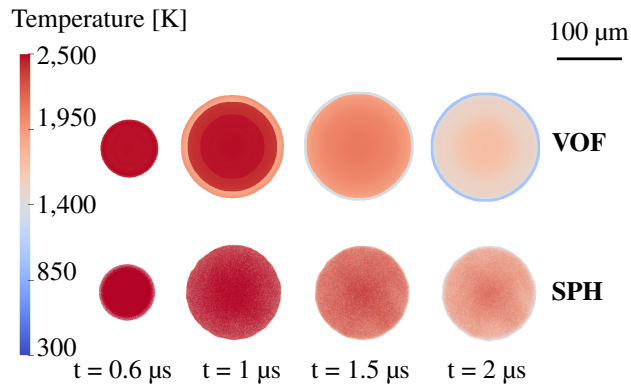}
\caption{Top view of the simulated splats in SPH (bottom) and Ansys Fluent/VOF=0.5 (top) for several times. Times $t$ = \SI{1.5}{\micro\second} and  $t$ = \SI{2.0}{\micro\second} for Ansys Fluent are included for reference, although the resolution of the mesh is too low for the splat thickness to derive meaningful results.}
\label{fig:Draufsicht}
\end{figure}

Figure \ref{fig:Draufsicht} presents the top view for the droplet impact for several points in time. 
It can be seen that the shape of the splats in Fluent and SPH show an almost perfectly symmetric shape. 
Furthermore, it was observed that the splat seems to cool off faster in Fluent than in SPH. 
However, it should be noted here again, that the visible surface in Ansys Fluent corresponds to the volume fraction 0.5 of the ceramic phase. 
Additionally, for times $t \geq$ \SI{1.0}{\micro\second} the thickness of the splat became smaller than 5 to 6 mesh cells (see also Figure \ref{fig:vof_cross_section}), which is problematic in the VOF approach, as it disperses the boundary of the free surface and therefore requires several mesh cells for the transition from one liquid to the other. 
When the ratio of the number of cells in the transition region to the number of cells with volume fraction 1.0 becomes very large, it becomes more difficult to make accurate observations about the enclosed volume.
This is due to the fact that observations about the enclosed volume become highly sensitive to the selection of the contoured volume fraction.
It is therefore concluded that the results for $t \geq$ \SI{1.0}{\micro\second} should be considered with care, but they are included in this figure for reference.

\begin{figure}[t]
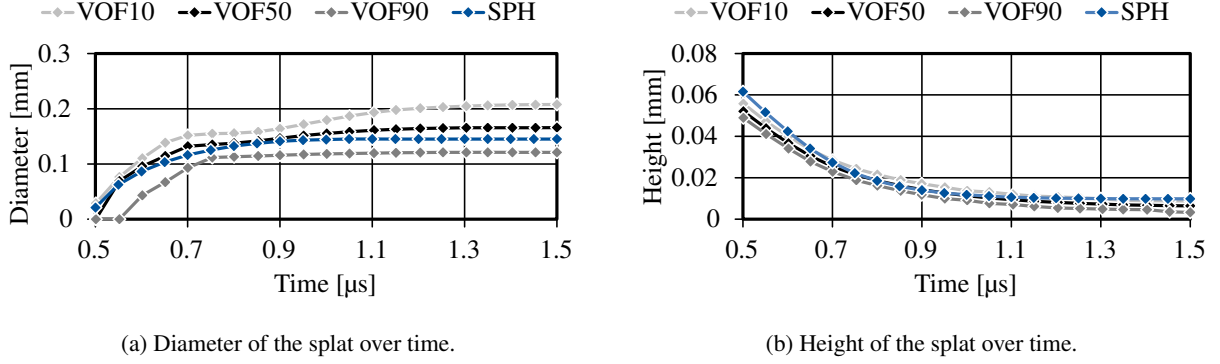

  \centering
\begin{subfigure}{.49\linewidth}
\def\svgwidth{1\textwidth}
  \input{Bilder_Aufprall_Fluent/Diameter_v2.eps_tex}
\caption{Diameter of the splat over time.}
\label{fig:diameter_splat}       
\end{subfigure}\hfill
\begin{subfigure}{.49\linewidth}
\def\svgwidth{1\textwidth}
  \input{Bilder_Aufprall_Fluent/Height.eps_tex}
\caption{Height of the splat over time.}
\label{fig:height_splat}       
\end{subfigure}
\caption{Comparison of the diameter and height of the simulated splats over time for SPH and Ansys Fluent at several volume fractions (0.1, 0.5, 0.9) of the ceramic phase.}
\label{fig:diameter_height_splat_v2}
\end{figure}

The diameter and the height of the formed splat were compared for SPH and Ansys Fluent for several volume fractions of the ceramic phase (0.1, 0.5, 0.9) over time in Figure \ref{fig:diameter_height_splat_v2}. 
The droplet impacted the substrate at \SI{0.5}{\micro\second} and subsequently spread out, gaining in diameter and losing in height until it reached a steady state. 
When examining the diameter and height of the splat calculated with the VOF method in Ansys Fluent, it was found that the dimensions vary significantly depending on the volume fraction, which is in accordance with the observation made in Figure \ref{fig:vof_cross_section}. 
It can be seen in Figure \ref{fig:diameter_splat} that there is a good agreement in diameter for SPH compared with the VOF method. The diameter calculated with SPH lies between that of volume fraction 0.5 and 0.9. It should be noted that the parameter of the cohesion correction, see Section \ref{sec:correction}, was adjusted manually to reach this agreement.
As discussed in the analysis of Figure \ref{fig:Draufsicht}, the presented results of Ansys Fluent for $t \geq$ \SI{1.0}{\micro\second} do not have a sufficient mesh resolution, despite having a total of 330k cells, and are therefore not discussed further. 

The height, as seen in Figure \ref{fig:height_splat} shows the same tendency for both approaches but a consistently smaller height was observed for all volume fractions of the ceramic phase in Ansys Fluent when compared to SPH. 
This is again consistent with the observed decrease in cross-section area in Figure \ref{fig:vof_cross_section} and in mass share of the droplet in Figure \ref{fig:volume_loss}. However, since both the height as well as the diameter have a strong influence on the heat transfer from the splat to the substrate, this difference is of high significance.
We attribute higher confidence to the SPH result regarding the height, because it does not suffer from the aforementioned volume dispersion.

\begin{figure}[t]
  \centering
\begin{subfigure}{.49\linewidth}
\def\svgwidth{1\textwidth}
  \input{Bilder_Aufprall_Fluent/vx_v2.eps_tex}
\caption{Comparison of the maximum, minimum and average radial velocity in Ansys Fluent and SPH, taken over the half space of a symmetry plane.}
\label{fig:radial_velo}      
\end{subfigure}\hfill
\begin{subfigure}{.49\linewidth}
\def\svgwidth{1\textwidth}
  \input{Bilder_Aufprall_Fluent/vy_v2.eps_tex}
\caption{Comparison of the maximum, minimum and average axi-al velocity in Ansys Fluent and SPH, the in-flight velocity of the droplet is directed in negative y-direction.}
\label{fig:axial_velo}       
\end{subfigure}
\caption{Comparison of the maximum, minimum and average radial and axial velocity of the droplet.}
\label{fig:velocity_droplet_v2}
\end{figure}

Figure \ref{fig:velocity_droplet_v2} shows a comparison of the simulated maximum, minimum, and average velocity over time taken over a half space in radial (x) and height (y) direction of 
Ansys Fluent (at VOF$=0.5$) and SPH. 
As the droplet impacts the substrate at $t=$ \SI{0.5}{\micro\second}, it can be seen in Figure \ref{fig:radial_velo} that shortly after impact a very strong increase of the maximum radial velocity from the initial maximum radial velocity of \SI{0}{\meter\per\second} occurs for both methods. 
This increase reaches roughly \SI{350}{\meter\per\second} for SPH, while the maximum radial velocity reaches nearly \SI{600}{\meter\per\second} in the case of Ansys Fluent. 
After this initial increase, the maximum radial velocity decreases towards zero for both cases for the time frame considered. 
Compared to this, the average velocity taken over a half space of both cases shows good agreement, with SPH exhibiting an consistently lower average radial velocity of the whole time frame considered.
Furthermore, it can be observed that the minimum velocity in the Ansys Fluent case remains zero for the entire duration, while the minimum velocity in SPH becomes negative, with a small but distinct minimum of approximately \SI{-40}{\meter\per\second} at the moment of impact. 
During the time shortly after impact, the negative velocities in radial direction are somewhat contrary to the expected dynamic of the process, in which the fluid of the droplet would spread outward (positive velocity in radial direction) to form the splat.
Upon further investigations, these particles are generally located near or on the cut plane where due to particle disorder some particles accelerating in negative x direction may appear.

In Figure \ref{fig:axial_velo} the maximum, minimum, and average velocity in vertical direction are shown. 
Please note, that the droplet moves towards the substrate, i.e. in negative y-direction. 
After impact at $t$ = \SI{0.5}{\micro\second}, the maximum vertical velocity decays smoothly to zero in the case of Ansys Fluent. 
In contrast to this, the observed maximum vertical velocity shows a slightly different behavior in SPH. 
It has a peak of almost -400 \si{\meter\per\second} at the time of impact $t$ = \SI{0.5}{\micro\second} before decreasing smooth\-ly towards zero.
The maximal vertical velocities of both cases for $t \geq$ \SI{0.6}{\micro\second} show an otherwise excellent agreement.
Similarly, the minimum velocity drops rapidly after impact and remains at zero or close to zero in the case of Ansys Fluent, while the minimum velocity simulated in SPH shows a different post-impact behavior.
At the time of the impact, the minimum velocity reaches an absolute minimum of roughly \SI{50}{\meter\per\second} before approaching zero.

\begin{figure}
  \centering
\def\svgwidth{.5\linewidth}
  \input{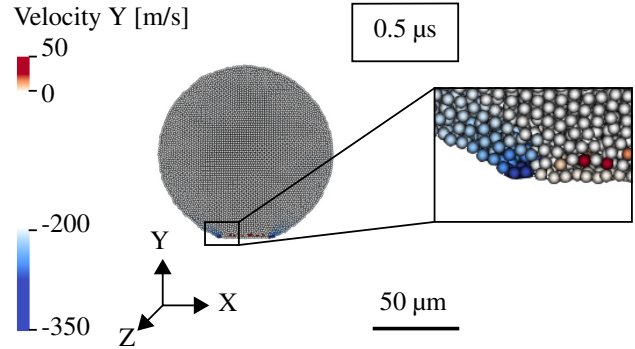}
\caption{Axial velocity at the moment of impact; the in-flight velocity of the droplet is directed in negative y-direction.}
\label{fig:sprung_geschwindigkeit}
\end{figure}

It is noticeable that the peak of the maximum velocity precedes the negative peak of the minimum velocity. While this can be understood in terms of a rebound effect, the reason for the large spread between minimum and maximum velocity, as well as the deviation with Ansys Fluent of these observables will be discussed later in detail in the analysis of Figure \ref{fig:sprung_geschwindigkeit}.
Finally, the average velocities of both simulation methods have an excellent agreement over time, even better than was the case for the radial velocity in Figure \ref{fig:radial_velo}. The droplet starts with a velocity of \SI{-200}{\meter\per\second} in both methods, then the average velocity in vertical direction decreases gradually after impact and reaches zero at $t$ = \SI{1.0}{\micro\second}.

A more detailed analysis of the apparent disagreement noted in Fig. \ref{fig:axial_velo} can be seen in Figure \ref{fig:sprung_geschwindigkeit} for the moment of impact at $t$ = \SI{0.5}{\micro\second}.
It can be seen in Figure \ref{fig:sprung_geschwindigkeit}, that a small fraction of particles at the side of the droplet have a very high vertical velocity of \SI{-350}{\meter\per\second} (color-coded in blue). 
Next to particles that are in contact with the wall, a small fraction of particles experience the rebound effect and their velocity reach nearly \SI{50}{\meter\per\second} (color-coded in red), before being counteracted by the bulk movement of the droplet. The main bulk of the particles has a velocity range from \SI{0} to \SI{-200}{\meter\per\second}, which corresponds to the the in-flight velocity of the droplet and solidified particles.
This observed peak of the velocity at the moment of impact, is assumed to be the result of the sudden difference in velocity due to solidification of the fluid and the subsequent increase of local density and jump in pressure. 
However, this does not necessarily imply an un-physical result, but on the contrary it might actually capture the real conditions even more accurately than the Eulerian method in Ansys Fluent.

Note that the investigation of minimum and maximum velocities are difficult to compare quantitatively and are only discussed in order to give better insight into the dynamics of the process for each simulation method.
While the actual values differ slightly, the overall trends visible in the minimum and maximum velocities are very similar for both methods.

\section{Discussion}
\label{sec:discussion}
In the previous section we compared the results of our SPH simulation against a simulation using the commercial tool Ansys Fluent.

While in the real process, of course, the partial melting of the droplet and the heat transfer cannot be neglected, including the latent heat of melting and solidification, in this work the aim was to compare the performance of the fundamentally different numerical methods at the conditions present at the droplet impact in thermal spraying. 
It is therefore considered justified to keep the model as simple as possible, also neglecting most non-linearities like temperature dependent material parameters for the comparison, although their implementation is possible.

We were able to use identical physical models and parameters for all phenomena, except for the corrective terms employed to improve the accuracy of the SPH simulation.
The corrective factors were adjusted to be as small as possible in order to closely match the droplet dimensions computed by Ansys Fluent.
From this we were able to obtain excellent results and good agreement with the Ansys Fluent simulation.

In terms of computational efficiency our proposed SPH method also compares very favorably to Ansys Fluent.
In our SPH method we used a total of 230k particles, while Ansys Fluent used a total number of 330k mesh cells.
While it may seem at first that the discretization using SPH is coarser, the actual discretization density in the region of interest, i.e. in the droplet, was $\sim 2.8^3 \approx 22$ times higher than the discretization density used by Ansys Fluent, whose mesh edge-length was \SI{2.25}{\micro\meter} which equated to 2.8 times the particle diameter of \SI{0.8}{\micro\meter}.
At the same time, our SPH method was able to finish the simulation in roughly 5 minutes, while Ansys Fluent required roughly 20 minutes.
This is a remarkable result, as it allows both a significant refinement in the region of interest, while reducing the simulation time by factor of 4.
All the more interesting are the scaling implications for the simulation of multiple droplet impacts. 
The increased performance allows for faster iteration times when simulating multiple droplets as well as significantly more accurate resolution of gaps in the coating on the scale of $L_{M}< \SI{1}{\micro\meter}$, than was possible before by Bobzin et al.~\cite{BWH21}.

While a large part of the SPH code is already well optimized, there are also some simple optimizations in reach that could further improve the SPH simulation performance.

\begin{figure*}
\centering
  \includegraphics[width=.8\textwidth]{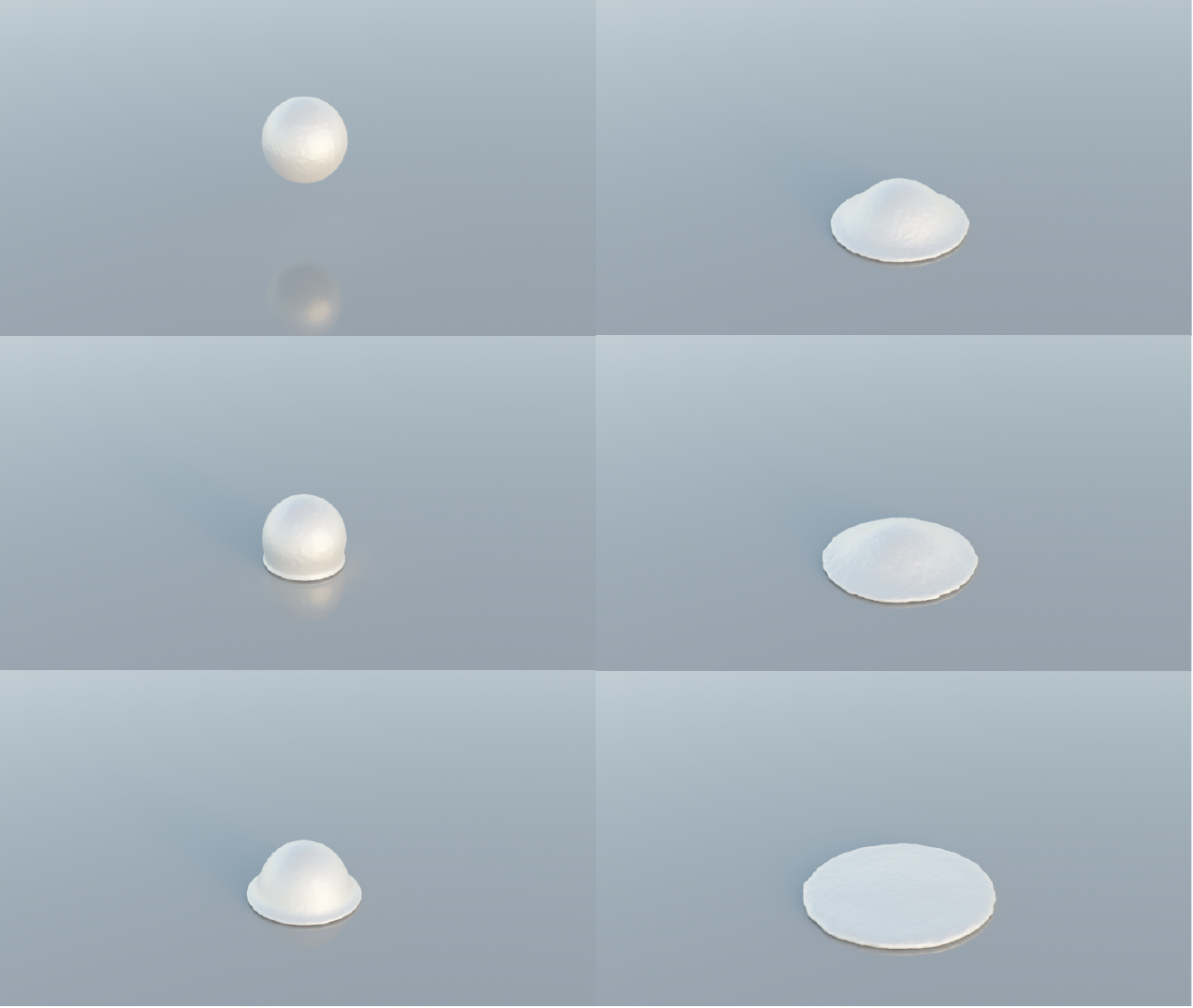}
\caption{Ray-traced rendering of the droplet impact dynamic simulated with SPH.}
\label{fig:tropfenaufprall}
\end{figure*}

Finally, we have shown in our evaluations that the selection of the liquid fraction for contouring of the ceramic phase has a very significant impact on droplet diameter, but especially the droplet height.
This has to do with the dispersion of the liquid surface that is ever-present for FVM-simulations using the VOF approach and adds an additional restriction on the mesh resolution.
Using our SPH method we were able to completely avoid the issue of dispersion and obtain a high-resolution simulation with a clearly defined surface.
A ray-traced rendering of a surface reconstruction of the SPH particle data is shown in Figure \ref{fig:tropfenaufprall}.

\section{Conclusion}
\label{sec:conclusion}
In this article we have shown that it is possible to simulate a molten droplet impact of the thermal spray process using nearly identical physical parameters in an SPH discretization as well as an FVM (finite volume method) discretization in Ansys Fluent.
We were able to perform a quantitative analysis of the simulations by considering droplet height, diameter and velocity distribution over time.
All of these showed good agreements, while the few dissimilarities were isolated and explained.

We introduced a novel SPH model which uses implicit integration for all forces except surface tension.
Because of this, our simulations remain stable for a wide range of large time steps.
We were also able to show that our SPH method is a very efficient and accurate alternative to the commercial FVM method of Ansys Fluent.
Our SPH method is able to have a higher discretization density in the region of interest while only requiring a quarter of the simulation time.

As a next step, we can build upon this work by considering multiple droplets of varying size and velocity, temperature dependent material parameters and heat transfer in the substrate.
When considering multiple droplets, the gaps in the coating may also be evaluated to a higher degree of accuracy than was previously possible.
A further extension could enable the simulation of multiple, only partially melted droplets of with both varying size as well as varying ratio of solid material at the core.
Furthermore, the consideration of rough surfaces and resulting contact angles could also be interesting for future work.


\section*{Acknowledgements}

The presented investigations were carried out at RWTH Aachen University within the framework of the Collaborative Research Centre SFB1120-236616214 “Bau\-teilpräzision durch Beherrschung von Schmelze und Erstarrung in Produktionsprozessen” and funded by the Deutsche Forschungsgemeinschaft e.V. (DFG, German Research Foundation). The sponsorship and support is gratefully acknowledged. Simulations were performed with computing resources granted by RWTH Aachen University under project rwth0570.

%
\section*{Conflict of interest}

On behalf of all authors, the corresponding author states that there is no conflict of interest.

\bibliographystyle{spmpsci}      

\bibliography{vof_cleaned.bib}   

\end{document}